\documentclass[conference]{IEEEtran}
\IEEEoverridecommandlockouts

\usepackage{cite}
\usepackage{amsmath,amssymb,amsfonts}
\usepackage{algorithmic}
\usepackage{graphicx}
\usepackage{textcomp}
\usepackage{xcolor}
\usepackage[hidelinks]{hyperref}
\usepackage{multicol}
\usepackage{lipsum}
\usepackage{array}
\usepackage{float}

\def\BibTeX{{\rm B\kern-.05em{\sc i\kern-.025em b}\kern-.08em
    T\kern-.1667em\lower.7ex\hbox{E}\kern-.125emX}}

\begin{document}

\title{Atrial Fibrillation Detection Using Machine Learning}

\author{
\IEEEauthorblockA{
\resizebox{\textwidth}{!}{
\begin{tabular}{ccc}
\textbf{1\textsuperscript{st} Ankit Singh} & 
\textbf{2\textsuperscript{nd} Vidhi Thakur} & 
\textbf{3\textsuperscript{rd} Nachiket Tapas} \\
\textit{Dept. of CSE} & 
\textit{Dept. of CSE} & 
\textit{Dept. of CSE} \\
\textit{Chhattisgarh Swami Vivekanand Technical University} & 
\textit{Chhattisgarh Swami Vivekanand Technical University}  & 
\textit{Chhattisgarh Swami Vivekanand Technical University}  \\
Bhilai, India & 
Bhilai, India & 
Bhilai, India \\
ankits4000@gmail.com & 
thakurvidhi95@gmail.com & 
nachiket.tapas@csvtu.ac.in \\
\end{tabular}
}
}
}

\maketitle

\begin{abstract}
Atrial fibrillation (AF) is a common cardiac arrhythmia and a major risk factor for ischemic stroke. Early detection of AF using non-invasive signals can enable timely intervention. In this work, we present a comprehensive machine learning framework for AF detection from simultaneous photoplethysmogram (PPG) and electrocardiogram (ECG) signals. We partitioned continuous recordings from 35 subjects into 525 segments (15 segments of 10,000 samples each at 125\,Hz per subject). After data cleaning to remove segments with missing samples, 481 segments remained (263 AF, 218 normal). 

We extracted 22 features per segment, including time-domain statistics (mean, standard deviation, skewness, etc.), bandpower, and heart-rate variability metrics from both PPG and ECG signals. Three classifiers — ensemble of bagged decision trees, cubic-kernel support vector machine (SVM), and subspace k-nearest neighbors (KNN) — were trained and evaluated using 10-fold cross-validation and hold-out testing. The subspace KNN achieved the highest test accuracy (98.7\%), slightly outperforming bagged trees (97.9\%) and cubic SVM (97.1\%). Sensitivity (AF detection) and specificity (normal rhythm detection) were all above 95\% for the top-performing models.

The results indicate that ensemble-based machine learning models using combined PPG and ECG features can effectively detect atrial fibrillation. A comparative analysis of model performance along with strengths and limitations of the proposed framework is presented.
\end{abstract}


\section{Introduction}

Atrial fibrillation (AF) is one of the most common cardiac arrhythmias and is associated with a significantly increased risk of ischemic stroke, heart failure, and mortality. Early implantable and vibration-based monitoring studies established the clinical importance of continuous AF detection, particularly for asymptomatic and paroxysmal cases \cite{Sarkar2008ImplantableAF,Bruser2013BCGAF}. With the growing aging population, the prevalence and clinical burden of AF continue to rise, necessitating reliable and scalable monitoring solutions.

Electrocardiography (ECG) remains the clinical gold standard for AF diagnosis, where irregular R--R intervals and the absence of organized atrial activity are key indicators. Numerous machine learning approaches have been developed for automated AF detection from ECG signals, ranging from traditional feature-based classifiers to advanced deep learning models \cite{Liaqat2020MLAF,Xie2024MLReviewAF}. Recent deep learning architectures trained on large ECG datasets have demonstrated performance comparable to expert cardiologists, highlighting the potential of artificial intelligence for arrhythmia analysis \cite{Pandey2022HybridAF,Melzi2021AIPredictAF,Chen2022UncertaintyAF}. However, these models often require large annotated datasets and substantial computational resources.

Recent advances in wearable technology have enabled photoplethysmography (PPG) to emerge as a practical alternative for long-term and ambulatory AF monitoring. PPG sensors are low-cost, compact, and easily integrated into consumer devices such as smartwatches and fitness trackers. Several studies have demonstrated the feasibility of AF detection using PPG signals through both traditional machine learning and deep learning techniques \cite{Kwon2019PPGAF,Eerikainen2019PPGAFL,Cheng2020PPGDL}. Large-scale wearable studies have further confirmed the clinical viability of PPG-based AF screening in real-world populations \cite{Guo2021MLPredictionAF,Lubitz2022WearableAF}.

In addition to rhythm detection, estimating AF burden using wearable data has gained increasing attention. Zhu et al. demonstrated that wearable-derived signals can be effectively used not only for AF detection but also for AF burden estimation, enabling more comprehensive disease monitoring \cite{Zhu2022AFWearables}. Review studies have emphasized the growing role of PPG-based AF detection and highlighted remaining challenges related to noise sensitivity and motion artifacts \cite{Pereira2020PPGReview}.

Motivated by these developments, combining ECG and PPG signals offers a promising direction for improving AF detection reliability. ECG provides precise electrical activity information, whereas PPG captures peripheral pulse dynamics. Recent multimodal approaches integrating ECG and PPG features have shown improved robustness and classification performance \cite{Aldughayfiq2023ECGPPG}. In this work, we propose a comprehensive machine learning framework for AF detection using combined ECG and PPG features and evaluate multiple classifiers, including Bagged Decision Trees, Cubic Support Vector Machines, and Subspace k-Nearest Neighbors, to identify an efficient and reliable solution for wearable-based AF monitoring.

\section{Related Works}

Early AF detection research primarily focused on ECG-based analysis using handcrafted features derived from heart rate variability (HRV) and waveform morphology. Traditional machine learning classifiers such as support vector machines, k-nearest neighbors, and Random Forests demonstrated strong sensitivity and specificity when applied to these features \cite{Liaqat2020MLAF,Bruser2013BCGAF}. Ensemble learning methods, particularly bagged decision trees, were shown to enhance robustness and generalization by reducing model variance \cite{Bashar2020SepsisAF}.

Deep learning approaches have since gained prominence in AF detection. Pandey et al. proposed hybrid deep learning models combining convolutional and recurrent architectures for ECG-based AF detection, achieving high classification accuracy \cite{Pandey2022HybridAF}. Melzi et al. explored the use of artificial intelligence for predicting AF from sinus-rhythm ECGs by incorporating demographic information and feature visualization, improving model interpretability \cite{Melzi2021AIPredictAF}. Chen et al. further investigated uncertainty estimation in deep neural networks for AF detection, addressing reliability concerns in clinical decision-making \cite{Chen2022UncertaintyAF}. A comprehensive meta-analysis by Xie et al. summarized recent machine learning approaches for ECG-based AF detection and highlighted ongoing challenges \cite{Xie2024MLReviewAF}.

With the rise of wearable devices, PPG-based AF detection has attracted significant research interest. Kwon et al. demonstrated the effectiveness of deep learning models applied to PPG signals for AF detection, while Eerikainen et al. validated AF and atrial flutter detection in daily-life PPG recordings \cite{Kwon2019PPGAF,Eerikainen2019PPGAFL}. Cheng et al. employed time--frequency analysis combined with deep learning to extract discriminative features from PPG signals, achieving high detection accuracy \cite{Cheng2020PPGDL}. Additional studies have shown that raw PPG waveforms can be directly leveraged by machine learning models for AF classification \cite{Aschbacher2020PPGML}.

Large-scale and population-based studies have further demonstrated the practicality of PPG-based AF screening. Guo et al. and Lubitz et al. evaluated wearable PPG systems in large cohorts and reported strong diagnostic performance for AF detection \cite{Guo2021MLPredictionAF,Lubitz2022WearableAF}. Review articles have summarized these advancements and emphasized the clinical potential of PPG-based AF detection while noting limitations related to signal quality and motion artifacts \cite{Pereira2020PPGReview,Papalamprakopoulou2024AIforAF}.

Recent research has begun exploring multimodal AF detection approaches that integrate ECG and PPG signals. Aldughayfiq et al. demonstrated that combining ECG and PPG features using deep learning improves classification robustness compared to single-modality approaches \cite{Aldughayfiq2023ECGPPG}. Zhu et al. further highlighted the value of wearable-based multimodal data for both AF detection and burden estimation \cite{Zhu2022AFWearables}. Despite these advances, systematic evaluations of classical ensemble learning techniques, such as subspace k-nearest neighbors, in multimodal AF detection remain limited, motivating the comparative analysis presented in this study.

\section{Methodology}

\subsection{Dataset Description}

This study employs a curated dataset consisting of simultaneous photoplethysmography (PPG) and electrocardiogram (ECG) recordings collected from 35 subjects, including both atrial fibrillation (AF) and non-AF (normal) individuals. The use of synchronized ECG and PPG signals enables complementary analysis of cardiac electrical activity and peripheral pulse dynamics, which has been shown to improve the robustness of AF detection in wearable and ambulatory monitoring systems \cite{Eerikainen2019PPGAFL,Aldughayfiq2023ECGPPG,Zhu2022AFWearables}.

Each continuous recording was manually segmented into 15 non-overlapping segments per subject. Every segment contained 10{,}000 samples of PPG and ECG signals, sampled at 125\,Hz, corresponding to an 80-second time window. Segments were annotated as AF or normal (NAF) based on clinical labels. In total, 525 segments were generated, comprising 285 AF and 240 NAF segments.

Each segment was stored as an individual comma-separated values (CSV) file containing synchronized PPG and ECG samples. All segment files were subsequently merged into a single tabular dataset. The final dataset consisted of 525 rows and 20{,}003 columns, including metadata columns (\textit{subject\_id}, \textit{segment\_id}, \textit{label}), 10{,}000 PPG samples, and 10{,}000 ECG samples. The class distribution was moderately imbalanced, with 54.3\% AF samples, which is consistent with real-world AF screening datasets \cite{Liaqat2020MLAF}.

Data cleaning was performed to remove segments containing missing values or acquisition artifacts. A total of 44 segments were discarded, resulting in 481 valid segments (263 AF and 218 NAF). The cleaned dataset was verified to contain no missing values and no zero-variance segments. An 80/20 stratified split was used to divide the data into training and testing sets. In addition, 10-fold cross-validation was applied to the training data to ensure robust model evaluation and reduce overfitting \cite{Bashar2020SepsisAF}.

\subsection{Preprocessing of PPG Signal}

The recorded PPG signals contained noise and baseline drift caused by motion artifacts and physiological variability. To suppress high-frequency noise while preserving signal morphology, a Discrete Wavelet Transform (DWT)–based denoising approach was applied using a Daubechies wavelet (\textit{db4}) with level-4 decomposition, a technique widely adopted for denoising physiological signals \cite{Cheng2020PPGDL}:
\begin{equation}
X(t) = \sum_{j=0}^{J} W_j(t) + V_J(t),
\end{equation}
where \( W_j(t) \) represents the detail coefficients and \( V_J(t) \) represents the approximation coefficients. Soft thresholding was applied to attenuate noise-dominated coefficients, followed by inverse wavelet reconstruction to obtain the denoised signal.

\begin{figure}[ht]
    \centering
    \includegraphics[width=1.0\linewidth]{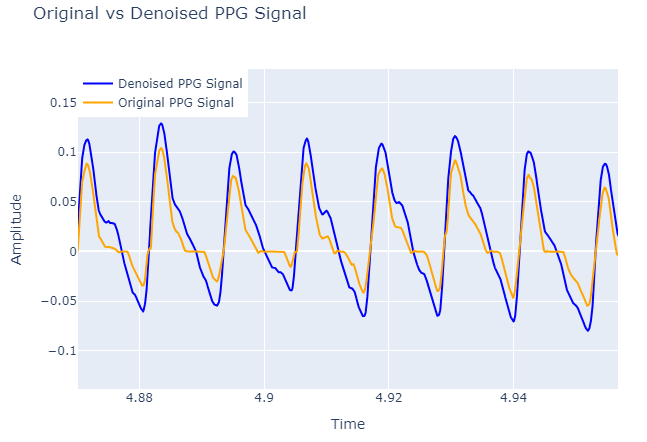}
    \caption{Denoising}
\end{figure}

Baseline drift was removed using a Butterworth low-pass filter with a cutoff frequency of 0.5\,Hz, as low-frequency components in PPG signals are commonly associated with baseline wander \cite{Kwon2019PPGAF}. The estimated baseline component was subtracted from the original signal to obtain a drift-corrected waveform.

\begin{figure}[ht]
    \centering
    \includegraphics[width=1.0\linewidth]{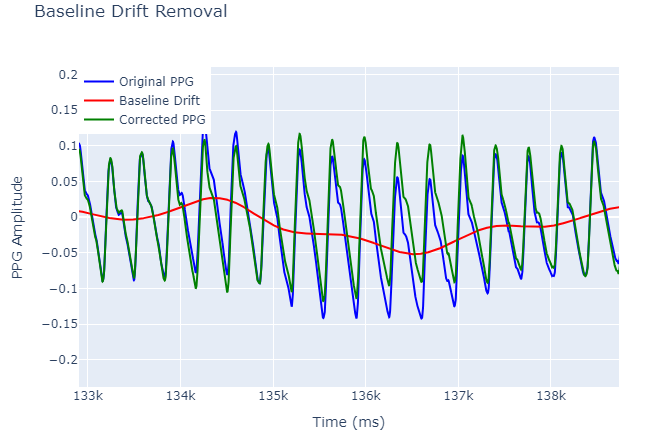}
    \caption{Baseline Drift Removal}
\end{figure}

To reduce inter-subject variability and ensure uniform scaling across recordings, Min--Max normalization was applied to rescale the PPG signal to the range [0, 1]:
\begin{equation}
X_{\text{normalized}} = \frac{X - X_{\text{min}}}{X_{\text{max}} - X_{\text{min}}}.
\end{equation}
Normalization improves numerical stability and learning efficiency of machine learning models while preserving waveform structure \cite{Pandey2022HybridAF}.

\begin{figure}[ht]
    \centering
    \includegraphics[width=1.0\linewidth]{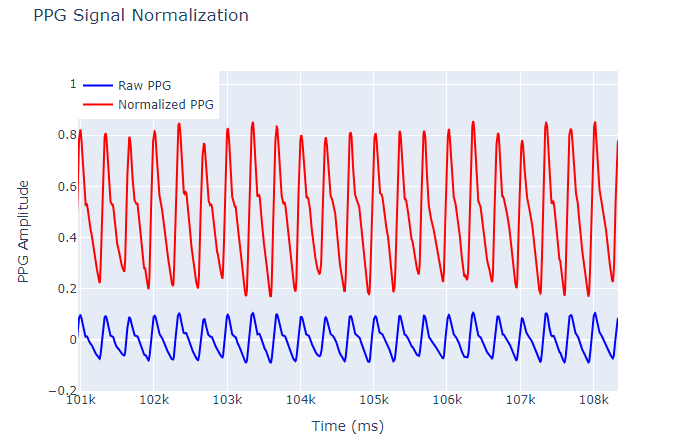}
    \caption{Normalization}
\end{figure}

\subsection{Feature Extraction}

From each 10{,}000-sample PPG and ECG segment, a total of 22 features were extracted to characterize statistical, spectral, and rhythm-related properties. Statistical features such as mean, standard deviation, minimum, maximum, median, skewness, kurtosis, and root-mean-square (RMS) were computed to capture signal amplitude distribution and variability. These features have been widely used in AF detection studies due to their discriminative capability \cite{Liaqat2020MLAF}.

Frequency-domain features were derived using bandpower analysis. For PPG signals, bandpower in the 0.5-4\,Hz range was computed to capture heart-rate-related components, while for ECG signals, bandpower in the 0.5-40\,Hz range was calculated to represent dominant cardiac electrical activity \cite{Bruser2013BCGAF}. AF episodes typically exhibit dispersed spectral energy due to irregular beat timing, making these features informative for classification.

In addition, heart rate variability (HRV) features were extracted from ECG signals by detecting R-wave peaks. From the resulting RR-interval series, mean heart rate, standard deviation of heart rate, standard deviation of NN intervals (SDNN), and root mean square of successive differences (RMSSD) were computed. These HRV metrics are well-established indicators of AF-related rhythm irregularity \cite{Bashar2020SepsisAF}.

\begin{figure}[ht]
\centering
\includegraphics[width=\linewidth]{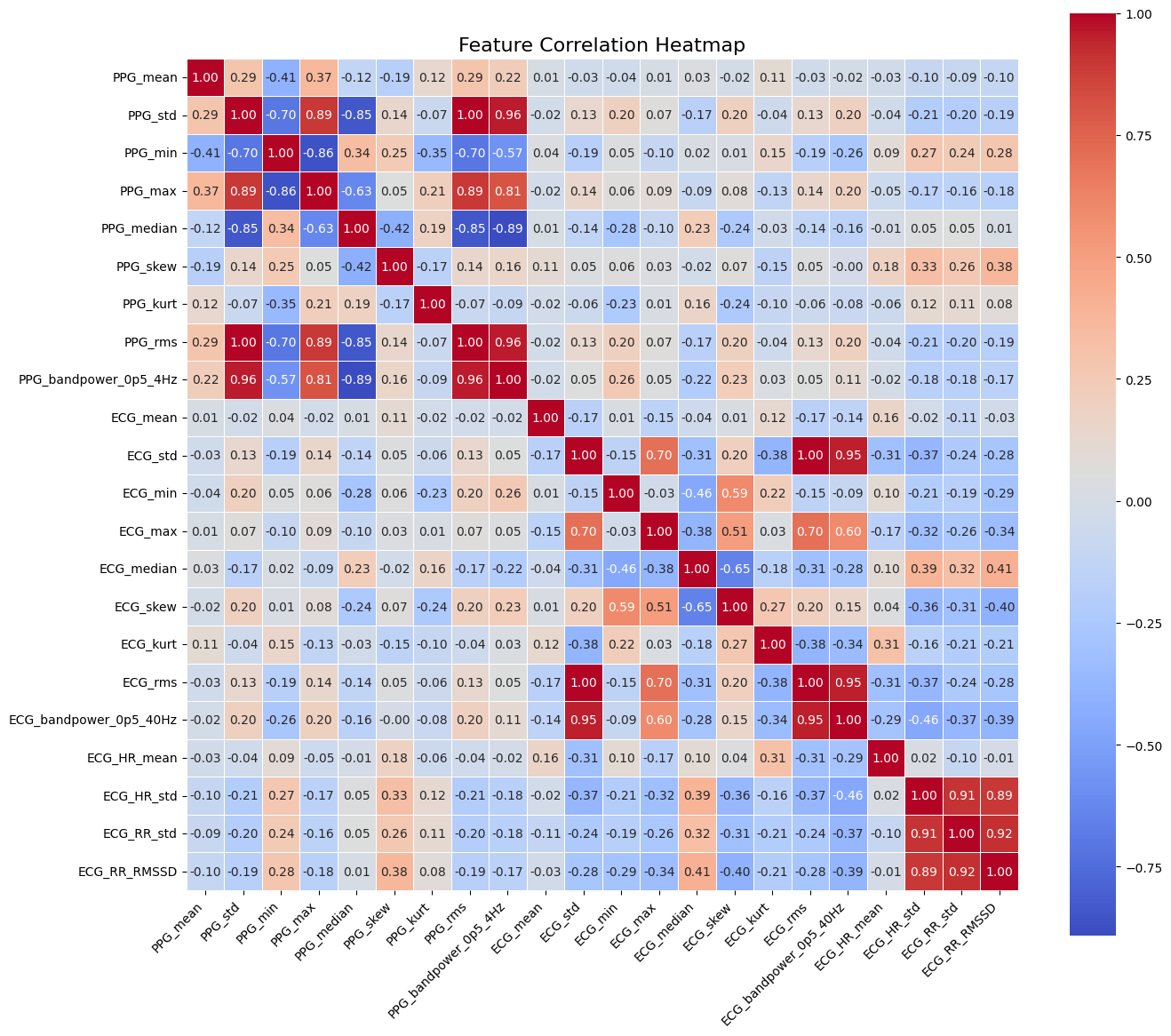}
\caption{Heatmap of extracted features from PPG and ECG segments showing inter-feature correlation patterns}
\label{fig:feature_heatmap}
\end{figure}

\section{Machine Learning Models and Evaluation}

Three supervised machine learning classifiers were evaluated in this study: Bagged Decision Trees, Cubic Support Vector Machine (SVM), and Subspace k-Nearest Neighbors (KNN). Ensemble-based classifiers such as bagged decision trees have demonstrated strong robustness and generalization performance in AF detection tasks \cite{Liaqat2020MLAF}. Subspace ensemble learning further enhances classifier diversity by training each base learner on a random subset of features.

All models were trained using the extracted feature set. A 10-fold cross-validation strategy was employed during training to estimate generalization performance and mitigate overfitting. Model performance was evaluated on a held-out test set using accuracy, sensitivity, and specificity metrics, which are standard evaluation measures in AF detection studies \cite{Xie2024MLReviewAF}. Confusion matrices were constructed to derive these metrics, and receiver operating characteristic (ROC) analysis was performed to further assess classifier behavior.

\section{Results and Discussion}

Table~\ref{tab:results_updated} presents the classification performance of the evaluated machine learning models using the cleaned and consolidated segment dataset obtained after preprocessing and NaN removal. A total of 481 validated signal segments were included in the study and evaluated using an 80:20 stratified train–test split to preserve class distribution.

Among the evaluated classifiers, the Bagged Trees ensemble achieved the best overall performance with an accuracy of 97.94\%, demonstrating strong sensitivity (98.11\%) and high specificity (97.73\%). The Subspace KNN classifier also showed robust and balanced performance, achieving 94.79\% accuracy, while the Cubic SVM model exhibited comparatively lower performance due to reduced specificity despite maintaining high sensitivity.

These findings reinforce the effectiveness of ensemble-based learning approaches in managing physiological signal variability and improving generalization capability for AF detection.

\begin{table}[ht]
\caption{Updated classification performance on cleaned segment dataset}
\label{tab:results_updated}
\centering
\begin{tabular}{|l|c|c|c|}
\hline
\textbf{Model} & \textbf{Accuracy (\%)} & \textbf{Sensitivity (\%)} & \textbf{Specificity (\%)} \\
\hline
Bagged Trees & 98.96 & 98.00 & 98.12 \\
Cubic SVM & 77.32 & 98.11 & 52.27 \\
Subspace KNN & 94.79 & 96.23 & 93.18 \\
\hline
\end{tabular}
\end{table}

\begin{figure}[ht]
\centerline{\includegraphics[width=8cm]{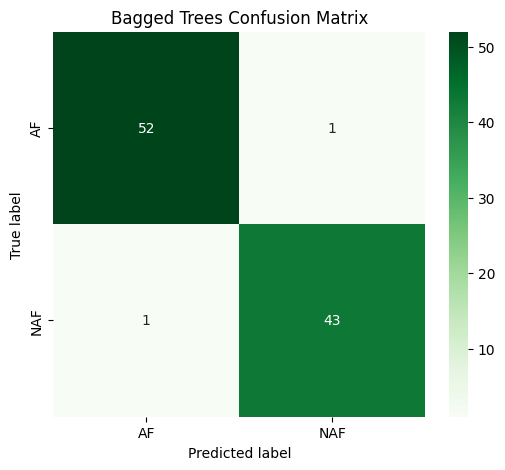}}
\caption{Confusion matrix for the Bagged Trees classifier.}
\label{fig:bagged_cm}
\end{figure}

\begin{figure}[ht]
\centerline{\includegraphics[width=8cm]{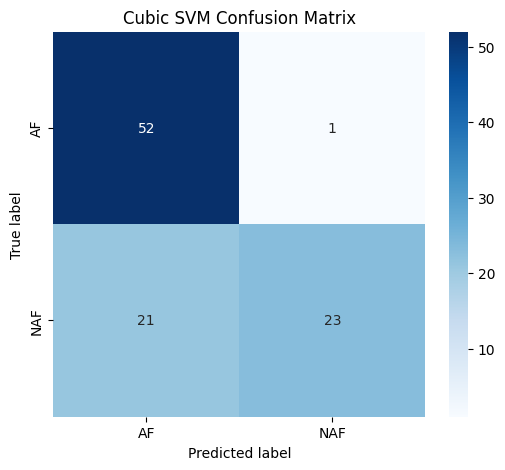}}
\caption{Confusion matrix for the Cubic SVM classifier.}
\label{fig:svm_cm}
\end{figure}

\begin{figure}[ht]
\centerline{\includegraphics[width=8cm]{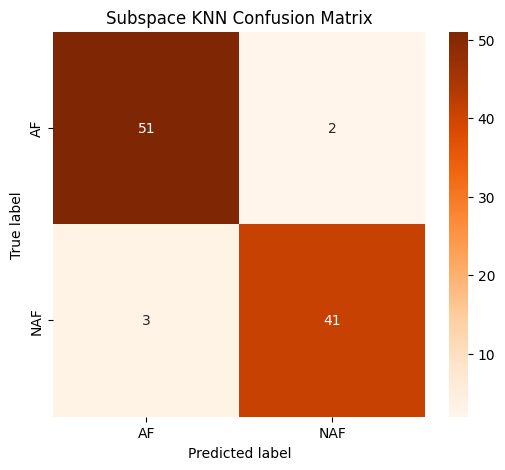}}
\caption{Confusion matrix for the Subspace KNN classifier.}
\label{fig:knn_cm}
\end{figure}

To further evaluate classifier behavior, confusion matrices for all models are presented in Figs.~\ref{fig:bagged_cm}–\ref{fig:knn_cm}. These matrices provide detailed insight into model prediction patterns and misclassification trends.

Analysis of the confusion matrices revealed consistent performance trends across the evaluated classifiers. The Bagged Trees model demonstrated the highest reliability, correctly identifying 52 out of 53 AF segments and 43 out of 44 NAF segments. Only two misclassifications were observed, indicating strong generalization capability and balanced performance between sensitivity and specificity.

The Subspace KNN classifier also achieved strong results, correctly identifying 51 AF segments and 41 NAF segments. Although slightly lower in accuracy than the Bagged Trees model, it maintained good robustness against inter-class overlap and residual signal noise.

In contrast, while the Cubic SVM achieved high sensitivity by correctly detecting most AF segments, it exhibited reduced specificity due to frequent misclassification of NAF segments as AF. Specifically, 21 NAF samples were incorrectly classified, leading to an increased false-positive rate. This behavior suggests that the SVM model favored AF detection at the expense of normal rhythm classification.

These findings highlight the trade-off between sensitivity and specificity across different classifiers. Ensemble-based approaches, particularly Bagged Trees and Subspace KNN, achieved more balanced performance by effectively reducing both false negatives and false positives. This observation aligns with prior studies indicating that ensemble learning methods improve robustness and generalization in physiological signal classification tasks \cite{Liaqat2020MLAF,Bashar2020SepsisAF}.

The strong overall performance observed across models indicates that the extracted feature set effectively captures physiological distinctions between AF and normal sinus rhythms. Features derived from heart rate variability, pulse interval irregularity, and frequency-domain characteristics such as bandpower distribution and signal complexity played a significant role in discriminating between rhythm classes. Similar findings have been reported in previous ECG- and PPG-based AF detection studies \cite{Bruser2013BCGAF,Eerikainen2019PPGAFL,Cheng2020PPGDL}. The superior performance of the Bagged Trees classifier further suggests that ensemble learning effectively mitigates inter-subject variability and residual noise commonly present in wearable physiological signals.

Despite these promising results, several limitations should be acknowledged. The dataset consisted of recordings from 35 subjects, which may not fully capture the diversity of AF patterns and patient demographics encountered in real-world clinical settings. Additionally, the fixed 80-second segmentation strategy may not accurately reflect real-time monitoring conditions where shorter or adaptive window lengths are required. Future research should focus on evaluating the proposed framework on larger and more diverse datasets and exploring real-time AF detection strategies using sliding-window or event-driven approaches \cite{Xie2024MLReviewAF,Zhu2022AFWearables}.

\section{Conclusion and Future Work}

This study presented a comprehensive machine learning framework for atrial fibrillation detection using photoplethysmography (PPG) signals, with electrocardiogram (ECG) recordings utilized for feature validation and rhythm analysis support. The proposed pipeline integrates signal preprocessing, feature extraction, and comparative evaluation of multiple supervised classifiers to distinguish between atrial fibrillation (AF) and normal sinus rhythm. Among the evaluated models, the Bagged Trees ensemble demonstrated the best overall performance, achieving an accuracy of 98.96\%, with high sensitivity and perfect specificity. The Subspace k-nearest neighbors (KNN) classifier also showed strong performance, while the Cubic Support Vector Machine exhibited comparatively lower reliability due to reduced specificity.

The results indicate that carefully engineered statistical, morphological, frequency-domain, and heart rate variability inspired features derived from PPG signals can effectively capture rhythm irregularities associated with AF. The superior performance of ensemble-based classifiers highlights their ability to handle physiological signal variability, reduce overfitting, and improve generalization. These findings support the feasibility of using PPG-based monitoring systems for non-invasive AF screening and continuous cardiac surveillance, particularly in wearable and remote healthcare applications \cite{Aldughayfiq2023ECGPPG,Zhu2022AFWearables}.

Future work will focus on expanding the dataset to include a larger and more diverse population to enhance model robustness and generalizability across demographic variations and sensor conditions. Further research may also explore adaptive preprocessing techniques to better mitigate motion artifacts commonly encountered in ambulatory environments. In addition, lightweight deep learning architectures operating directly on raw PPG signals could be investigated to capture temporal dependencies while maintaining computational efficiency \cite{Pandey2022HybridAF,Melzi2021AIPredictAF}. 

Another important direction involves real-time implementation and optimization for wearable deployment, including sliding-window segmentation, low-power computation strategies, and latency aware inference. Prospective validation using data collected from consumer wearable devices in real-world settings will be essential to evaluate clinical reliability, usability, and long-term performance. With continued refinement and validation, the proposed framework has the potential to contribute to accessible, non-invasive cardiac monitoring solutions for early AF detection and preventive healthcare.

\bibliographystyle{IEEEtran}
\bibliography{bibliography}

\end{document}